\documentstyle[12pt]{article}
\newcommand{\IR}{\hbox{{\rm I}\kern-.2em\hbox{\rm R}}}
\newcommand{\IC}{\hbox{{\rm I}\kern-.6em\hbox{\bf C}}}
\newcommand{\half}{{\scriptstyle\frac{1}{2}}}
\newcommand{\ov}{\overline}
\newcommand{\hf}{\hat{f}}
\begin{document}%

\title{\Large \bf
Berezin quantization\\
and unitary representations of Lie groups}
\author{
{\normalsize\sc D. Bar-Moshe$^*$ and M. S. Marinov$^{*,**}$ }\\
{\normalsize\it
$^*$Department of Physics, Technion-Israel Institute of Technology}\\
{\normalsize\it Haifa 32000, Israel}\\
{\normalsize\it
$^{**}$Center for Relativity, University of Texas at Austin}\\
{\normalsize\it Austin, TX 78712-1081, U. S. A.}\\
}

\date{(Contribution to Berezin Memorial volume)}

\maketitle

\begin{abstract}
In 1974, Berezin proposed a quantum theory for dynamical systems
having a K\"{a}hler manifold as their phase space. The system states
were represented by holomorphic functions on the manifold. For any
homogeneous K\"{a}hler manifold, the Lie algebra of its group of motions
may be represented either by holomorphic differential operators
(``quantum theory"), or by functions on the manifold with Poisson
brackets, generated by the K\"{a}hler structure (``classical theory").
The K\"{a}hler potentials and the corresponding Lie algebras are constructed
now explicitly for all unitary representations of any compact simple Lie group.
The quantum dynamics can be represented in terms of a phase-space path
integral, and the action principle appears in the semi-classical approximation.
\end{abstract}


\newpage
\section{Introduction}

\subsection{Historical background}

In a series of papers \cite{ber72,ber74,ber75a,ber75},
written in the beginning of the 70-ties, Berezin developed a new approach to
description of quantum dynamical systems on non-Euclidean
phase-space manifolds. The approach was essentially based upon two basic
implements: functional formalism and complex analysis.

{}From its very beginning, quantum mechanics
was expressed in terms of the Hilbert space of state vectors (represented, as
a rule, by wave functions in the coordinate representation) and observables
given by operators acting in that space. The fundamental observables,
like energy, momentum or angular momentum, were represented by differential
operators, so differential equations played the dominant role in the early
quantum theory. Fock was probably the first\cite{fock28,fock32} who
understood, still in the early days of quantum theory, that functional methods
may be extremely powerful, especially, in applications of the method of
 second quantization to quantum field theory. Schwinger's quantum action
principle and Feynman's path integrals were introduced into quantum
electrodynamics and did lead to its triumph in the late 40-ties. Berezin was
exposed to that brilliant development when he was a student in the 50-ties
and took part in the Landau seminars at the Institute of Physical Problems in
Moscow. At that time, there was evidently a missing element in the functional
representation of quantum electrodynamics: a proper description of fermion
fields. Schwinger employed {\em second-kind variables}\cite{schwin53}
in his action principle, in order to describe fermion fields, electrons in
particular, but the formalism of anti-commuting numbers was not yet
developed systematically. Berezin invented the integral in anti-commuting
(Grassmann) variables, and the unified functional approach to quantal systems
with Bose and Fermi degrees of freedom was presented in a complete form in his
thesis. His famous book\cite{book} was based upon the thesis, and it
has been the mathematical foundation for the future development of methods
of supersymmetry in physics\cite{bdewitt}. Later, spinning
string models prompted an elaboration of the functional
approach to particle spin dynamics and the Poisson brackets on
super-phase spaces were defined\cite{bermar75,bermar}.

The second quantization in its functional form is naturally related to complex
analysis. Fock\cite{fock32}, and later Bargmann\cite{barg62},
used the {\em holomorphic} representation of functionals in quantum theory.
(For reviews, see e.g. Refs.\cite{novo58,fad76}.) Investigation of
properties of normal and anti-normal symbols of the evolution operator
enabled Berezin to derive sophisticated theorems about spectra of Hamiltonian
operators\cite{ber72,ber72a} for the Euclidean phase space. Those ideas were
further developed later, e.g. in Ref.\cite{simon77}. On the other hand,
Berezin discovered that the formalism can be extended to
{\em non-Euclidean} dynamical systems, as soon as the corresponding
phase-space manifold has a complex K\"{a}hlerian structure.

\subsection{Quantization in Euclidean phase spaces}

Berezin's line of reasoning was fairly straightforward. It was known that in
the Euclidean case, the states of any quantal system can be represented by
holomorphic functions $\psi(z)$, where $z\equiv\{z_1,\cdots,z_m\}\in\IC^m$ and
$m$ is the number of degrees of freedom. The structure of the Hilbert space was
introduced in the space of the state vectors by means of the inner product
     \begin{equation}
(\psi_2,\psi_1)=\int_{\;\IC^m}\overline{\psi_2(z)}
\psi_1(z)e^{-(z\cdot\bar{z})}d\mu(z,\bar{z}).
   \end{equation}
Here $(z\cdot\bar{z})\equiv\sum^m_{\alpha=1}z_\alpha\bar{z}_\alpha$ is the
Euclidean scalar product, and the integral is taken with
the standard Liouville measure
     \begin{equation}
d\mu(z,\bar{z})\equiv
\prod^m_{\alpha=1}\frac{dz_\alpha\wedge d\bar{z}_\alpha}{-2\pi i}=
\prod^m_{\alpha=1}\frac{dq_\alpha dp_\alpha}{2\pi\hbar},\;\;\;
z=({\bf q}+i{\bf p})/\sqrt{2\hbar},
   \end{equation}
where $\bf q$ and $\bf p$ are the usual coordinate and momentum vectors.
The inner product is invariant under {\em translations} in the phase space,
if the wave functions have appropriate transformation properties,
     \begin{equation}
z\to z'=\epsilon z+a \leadsto
\psi(z)\to\psi'(z)=e^{\epsilon(z\cdot\bar{a})+\half (a\cdot\bar{a})}\psi(z'),
   \end{equation}
where $a$ is any complex $m$-vector and $|\epsilon|=1$.
After three consecutive translations by $a, b$ and $-a-b$, the phase space
remains invariable, while each wave function gets a constant phase shift
     \begin{equation}
\psi(z)\to\psi'(z)=e^{i\varphi}\psi(z),\;\;\;
\varphi=[(b\cdot\bar{a})-(a\cdot\bar{b})]/2i.
   \end{equation}
Evidently, $\half\varphi$ is just the Euclidean area of the triangle built upon
the vectors $a$ and $b$. In general, any continuous translation in the phase
space, given by a closed line in $\;\IC^m$, results in a phase shift, which is
the Euclidean area of the minimal surface stretched upon the line. Thus the
standard symplectic structure appears from Eq. (1), and one gets a geometric
interpretation for the canonical quantization: the space of state vectors is
a line bundle based upon the phase space. The phase space translations and
the phase shifts together constitute a Lie group ${\cal W}_m$
(Schwinger's {\em special canonical group}\cite{schwin70}).
The corresponding Lie algebra (the Heisenberg -- Weyl algebra) is given by
the canonical commutation relations for its basis elements, which are
the coordinate and momentum operators (or the creation and annihilation
operators) and the unit operator which generates the phase shifts (the
{\em gauge group} $U(1)\subset{\cal W}_m$ ).

The Berezin quantization stems from the extension of the holomorphic
quantization to phase spaces which are K\"{a}hler manifolds having non-flat
geometries. The inner product is defined by means of the integral with an
appropriate invariant measure, and the Hermitean scalar product in the
exponent is substituted for the K\"{a}hler potential, see Eq. (5) below.

\subsection{Contents of this paper}

The next Section presents a short review of Berezin's quantization principle.
A special attention is paid to the properties of cocycle functions which define
the structure of the line bundle representing the space of state vectors.
Two examples are also given: the two-dimensional sphere and the Lobachevsky
plane. Together with the usual phase plane, given in the previous subsection,
they represent all possible quantizations for one degree of freedom.
The quantization is given by a non-negative integer $l$, which
specifies the {\em quantized} curvature of the phase space.
In the compact case, i.e. for the sphere, the total
number of states of the system is finite, $N=l+1$. Section 3 shows a
correspondence between the quantization on a homogeneous  K\"{a}hler
manifold and the representation of the Lie algebra for its group of motion.
The Lie algebra is represented in two alternative ways:
i) by holomorphic differential operators, ii) by their symbols, i.e.
functions on the phase space and the appropriate Poisson brackets.
This gives the actual meaning to the word ``quantization".
In Section 4 the K\"{a}hler potentials are constructed explicitly for all
compact simple Lie groups. Complex coordinates correspond to positive
roots of the Lie algebra, and fundamental K\"{a}hler potentials are
expressed in terms of generalized determinants (considered in Appendix).
A description of
dynamics is discussed in Section 5. An analogue of the path integral is
derived, which leads to an action functional in the semi-classical
approximation, and the classical Hamilton equations of
motion result from the corresponding variational principle.

The paper is addressed mainly to physicists, and mathematical arguments
are more intuitive than rigorous.

\section{Quantization on homogeneous \hfill\break
K\"{a}hler manifolds}

\subsection{The Hilbert space of state vectors}

We shall consider a $m$-dimensional K\"{a}hler manifold $\cal M$
with local complex coordinates $z_\alpha$ ($\alpha=1,\cdots,m$),
where a group $\cal G$ is acting  holomorphically, i.e.
$z\to gz,\forall g\in {\cal G}$ (notations follow mainly
Ref.\cite{nomizu}). State vectors of the quantal system are defined
as holomorphic sections of the holomorphic line bundle $\cal L$ over
$\cal M$. The state vectors are represented by locally holomorphic
{\em wave functions} $\psi(z)$. The Hilbert space structure is assigned
to $\cal L$ by means of the following $\cal G$-invariant inner product,
     \begin{equation}
(\psi_2,\psi_1)=\int_{\cal M}\overline{\psi_2(z)}
\psi_1(z)\exp[-K(z,\bar{z})]d\mu(z,\bar{z}).
   \end{equation}
Here $K(z,\bar{z})$ is the K\"{a}hler potential, associated with the line
bundle $\cal L$ and defined in any open coordinate neighbourhood of the
manifold $\cal M$. The meaning of the integral includes, of course, the
sum over the neighbourhoods covering $\cal M$. The invariance is imposed
by the transformation law for the wave functions, which is consistent with
that for the K\"{a}hler potential, namely for any $g\in{\cal G}$
    \begin{eqnarray}
K(z,\bar{z})\to
K(gz,\overline{gz})=K(z,\bar{z})+\Phi(z;g)+\overline{\Phi(z;g)},\\
\psi(z)\to\psi(gz)=\exp [\Phi(z;g)](\hat{U}(g^{-1})\psi)(z),
\end{eqnarray}
where $\hat{U}(g)$ is a unitary operator representing the group element $g$
in the Hilbert space $\cal L$, $\hat{U}(g_1)\hat{U}(g_2)=\hat{U}(g_1g_2)$,
and the {\em cocycle function} $\Phi(z;g)$ is (locally) holomorphic.

The (invariant) integration measure is expressed, as usual, in terms of the
$m$-th power of the corresponding  K\"{a}hler (1,1)-form $\omega$,
             \begin{equation}
d\mu(z,\bar{z})\equiv C \frac{\omega}{2\pi i}
\wedge\cdots\wedge\frac{\omega}{2\pi i} \;(m\; {\rm times}),
       \end{equation}
where $C$ is a normalization constant (see Eq. (12) below) and
the factor $2\pi i$ is introduced for future convenience.
(We assume, of course, that the form $\omega$ is {\em non-degenerate},
so the integrals do not vanish identically.)
The (1,1)-form $\omega$ has the following local representation
     \begin{equation}
\omega\equiv\omega_{\alpha\bar{\beta}}(z,\bar{z})
dz^\alpha\wedge d\bar{z}^\beta;\;\;\;
\omega_{\alpha\bar{\beta}}=\partial_\alpha\partial_{\bar{\beta}}K,
  \end{equation}
where $\partial_\alpha\equiv\partial/\partial z^\alpha$,
$\partial_{\bar{\beta}}\equiv\partial/\partial \bar{z}^\beta$.
The form $\omega$ is closed,
$\partial\omega=0,$ $\bar{\partial}\omega=0,$
and invariant under the group transformations, as follows from Eq. (6).

Let us make two more assumptions.

(A) Group $\cal G$ acts transitively in $\cal M$, i.e. for any two points
there is a group element transforming one of them into the other.

(B) Excluding from $\cal M$ a manifold $\cal X$ of a lower dimensionality,
one gets a domain with a simple topology, ${\cal M\setminus X}\equiv\IC^m$.

It follows that any holomorphic function invariant under $\cal G$ is a
constant.
This property is very important for the following. Various manifolds $\cal X$,
to be excluded in order to reduce $\cal M$ to $\;\IC^m$, are transformed into
each other under the group transformations. Because of (B), the inner product
in Eq. (5) can be considered as the integral over $\;\IC^m$ and is independent
of the choice of $\cal X$.

\subsection{Symbols of linear operators}

For the manifolds of our concern here, the K\"{a}hler potential can be
considered as a boundary value of a function $K(\zeta,\bar{z})$,
holomorphic in the first variable and anti-holomorphic in the second one.
Berezin introduced a ``super-complete set" of state vectors
     \begin{equation}
\Psi_v(z)\equiv \exp[K(z,\bar{v})],
   \end{equation}
This set is an important class of {\em generalized coherent states}
(an exposition and an abundant bibliography can be found in Refs.
\cite{klauder,perelomov}).

One can prove that, under a proper normalization of the integration measure,
any element of the Hilbert space is reproduced by the integral on the manifold,
     \begin{equation}
\psi(\zeta)\equiv \int_{\cal M}\psi(z)
\exp\left[ K(\zeta,\bar{z})-K(z,\bar{z})\right]d\mu(z,\bar{z}),
\;\;\;\forall\psi\in{\cal L}.
    \end{equation}
The proof stems from the fact that, because of (6)-(7), the ratio of the
integral in the r.h.s. to $\psi(\zeta)$ is invariant under the group
transformations. This number is independent on $\psi$, and setting a proper
value of $C$ in Eq. (8), one can make it equal to 1. In order to prove that,
one can calculate $(\psi,\psi)$, applying (10) first to a coherent state and
then to $\psi$, getting an identity. Moreover, for the manifolds we are
dealing with, $K(0,\bar{z})\equiv 0$, and the following integral exists
     \begin{equation}
\int_{\cal M}\exp[-K(z,\bar{z})]d\mu(z,\bar{z})=1.
    \end{equation}
In the other words, all constant sections of $\cal L$ belong to the Hilbert
space, and one can assume that $\psi_0(z)\equiv 1$ has the unit norm.
(In typical physical problems a constant $\psi$ corresponds to the ``vacuum",
i.e. the system ground state for a proper Hamiltonian.) Thus the constant $C$
in Eq. (8) is expressed simply in terms of an integral over the manifold.

The reproducing kernel given in terms of the K\"{a}hler potential in Eq. (11),
has an expansion in terms of {\em any} orthonormal basis
$\{\phi_\nu (z)\}$ in $\cal L$,
     \begin{equation}
\exp[K(\zeta,\bar{z})]=\sum_\nu\phi_\nu(\zeta)\overline{\phi_\nu(z)}.
   \end{equation}
This equality gives also an expansion of the coherent state (10) in the
basis of the orthonormal states.

Linear operators in the Hilbert space, in particular describing
observables of the quantal system, are represented with their symbols
in the following way: $\hat{A}\rightarrow A(z,\bar{\zeta})$ means
     \begin{equation}
(\hat{A}\psi)(\zeta)=\int_{\cal M}A(\zeta,\bar{z})\psi(z)
\exp\left[ K(\zeta,\bar{z})-K(z,\bar{z})\right]d\mu(z,\bar{z}).
    \end{equation}
The symbol representation has the following nice properties.

1. The symbol of the unit operator is just 1,
$\hat{I}\rightarrow I(\zeta,\bar{z})\equiv 1$.

2. The trace of any operator is given by the integral of its symbol,
     \begin{equation}
{\rm tr}(\hat{A})= \int_{\cal M}A(z,\bar{z})d\mu(z,\bar{z}).
    \end{equation}
For compact manifolds, the trace of the unity operator $\hat{I}$ exists,
the volume is finite and equals the total number of states $N$,
     \begin{equation}
{\rm tr}(\hat{I})\equiv N=\int_{\cal M}d\mu(z,\bar{z}).
    \end{equation}

3. The Hermitean conjugation in the Hilbert space is represented by the
complex conjugation of the symbol and transposition of its arguments,
     \begin{equation}
\hat{A}^\dagger\rightarrow A^*(\zeta,\bar{z})=
\overline{A(z,\bar{\zeta})}.
    \end{equation}

4. The symbol for the product of operators is given by an integral of
the product of their symbols (the $*$-product)
     \begin{eqnarray}
\hat{A}\hat{B}\rightarrow (A*B)(\zeta,\bar{\eta})\equiv
\int_{\cal M}A(\zeta,\bar{z})B(z,\bar{\eta})\times\nonumber\\
\exp\left[ K(\zeta,\bar{z})-K(z,\bar{z})+K(z,\bar{\eta})-
K(\zeta,\bar{\eta})\right]d\mu(z,\bar{z}).
    \end{eqnarray}
In particular, one has an analogue of the Gaussian integral,
     \begin{equation}
\exp[K(\zeta,\bar{\zeta})]=\int_{\cal M}
\exp\left[
K(\zeta,\bar{z})-K(z,\bar{z})+K(z,\bar{\zeta})\right]d\mu(z,\bar{z}).
    \end{equation}
The scalar products of the coherent states (10) are obtained from (19).

In general, any system state is given by a (positive semi-definite) density
operator $\hat{\rho}$, which can be also represented with its symbol
$\rho(z,\bar{z})$.  For any operator $\hat{A}$, its expectation value in the
given state is
     \begin{eqnarray}
<A>_\rho\equiv{\rm tr}(\hat{A}\hat{\rho})=\int\!\!\int_{\cal M}
A(\zeta,\bar{z})\rho(z,\bar{\zeta})\times\\
\exp\left[ K(\zeta,\bar{z})-K(z,\bar{z})+K(z,\bar{\zeta})-
K(\zeta,\bar{\zeta})\right]d\mu(z,\bar{z})d\mu(\zeta,\bar{\zeta}).
\nonumber
    \end{eqnarray}
Thus the associative algebra of observables for the quantal system is
constructed completely in terms of the operator symbols.

\subsection{Cocycle functions}

The cocycle functions, defined in Eq. (6), have the following properties,
             \begin{eqnarray}
\Phi(z,e)=0,\;\;\;\;e-\;{\rm unity}\;\;{\rm in}\;{\cal G},\\
\Phi(gz;g^{-1})=-\Phi(z;g),\;\;\;\forall g\in{\cal G},\\
\Phi(z;g_2g_1)=\Phi(g_2z;g_1)+\Phi(z;g_2),\;\;\;\;
       \forall g_1,g_2\in {\cal G}.
          \end{eqnarray}
Writing the latter condition in the infinitesimal form, one gets differential
equations for the cocycle functions. In order to write them down, we need
some notations.

Let $\bf g$ be the Lie Algebra of the group $\cal G$. Introducing a basis
$\tau_a$ in $\bf g$,  with $a=1,\cdots, n\equiv {\rm dim}\;{\bf g}$, one has
Cartesian coordinates for the group elements
$g=\exp(-\xi^a\tau_a)$, and the corresponding (left) Lie derivatives $D_a$
on the group manifold. The action of the group $\cal G$ on the manifold
$\cal M$ determines the holomorphic Killing fields
    \begin{equation}
\nabla_a=\kappa^\alpha_a(z)\partial_\alpha,\;\;\;\;
\kappa^\alpha_a(z)\equiv D_a(gz)^\alpha\mid_{g=e}.
    \end{equation}
The conjugate Killing field and the differential operator $\ov {\nabla_a}$ are
defined similarly. The Lie derivative and the Killing derivative satisfy
the commutation relations of the Lie algebra $\bf g$,
     \begin{equation}
[\tau_a,\tau_b]=f^c_{ab}\tau_c\leadsto\;\;
[D_a,D_b]=f^c_{ab}D_c,\;\;\;[\nabla_a,\nabla_b]=f^c_{ab}\nabla_c.
\end{equation}
A holomorphic vector field is associated with the cocycle functions,
       \begin{equation}
\varphi_a(z)=D_a\Phi(z;g)\mid_{g=e}.
       \end{equation}
Now the differential equations for the cocycle functions are written down
as follows,
     \begin{equation}
D_a\Phi=\nabla_a\Phi+\varphi_a.   \end{equation}
The equations are consistent, as soon as
$\varphi_a(z)$ satisfy the linear differential equations
\begin{equation}
\nabla_a\varphi_b-\nabla_b\varphi_a=f^c_{ab}\varphi_c.
\end{equation}

In Eq. (6) the cocycle functions are defined up to an arbitrary imaginary term,
but this term determines the phase shift of the wave functions in Eq. (7).
As soon as the wave functions must be one-valued functions on $\cal M$,
the additional term must be a multiple of $2\pi i$, and one gets a boundary
condition on $\Phi$. Namely, let us consider a {\em compact} one-parameter
subgroup of $\cal G$, $u(t)\in U(1)\subset{\cal G}$, where $0\leq t<2\pi$.
For {\em any} such a subgroup, one must have
     \begin{equation}
\Phi(z;u(2\pi))=2\pi il,   \end{equation}
where $l$ is an integer, depending, in principle, on the equivalence class
of the subgroups $U(1)$. This is a {\em quantization condition} for the
K\"{a}hler manifolds, which is necessary for consistency of the quantum theory,
 based upon the line bundle structure, presented above. The quantization
condition is of exactly the same nature as the Dirac quantization for magnetic
charge\cite{diracm}, as explained by Wu and Yang\cite{wu}. If this condition
holds, the corresponding K\"{a}hler potential is called integral.
One can see that integral of $\omega$ over any two-dimensional cycle
stretched upon a line homotopical to the group trajectory is given by Eq. (29).

\subsection{Sphere and pseudosphere}

The simplest examples presented by Berezin\cite{ber75} for $m=1$
are the two-dimensional sphere $S^2$ and the pseudosphere $H^2$ (the
Lobachevsky plane). The following
table is a summary for these two manifolds,
     \begin{equation}
\begin{array}{rcc}
{\cal M}: & S^2 & H^2\\
{\cal G}: & SU(2) & SU(1,1)\\
K(z,\bar{z}): & l\log(1+z\bar{z}) & -(l+1)\log(1-z\bar{z}) \\
d\mu(z,\bar{z}): & (l+1)(1+z\bar{z})^{-2}dz\wedge d\bar{z}/2\pi i &
l(1-z\bar{z})^{-2}dz\wedge d\bar{z}/2\pi i \\
gz: & (\alpha z-\beta)/(\bar{\beta}z+\bar{\alpha}) &
(\alpha z+\beta)/(\bar{\beta}z+\bar{\alpha}) \\
\; & \alpha\bar{\alpha}+\beta\bar{\beta}=1 &
\alpha\bar{\alpha}-\beta\bar{\beta}=1\\
\Phi(z;g): & -l\log(\bar{\beta} z+\bar{\alpha}) &
(l+1)\log(\bar{\beta} z+\bar{\alpha})
\end{array}
   \end{equation}
In both the cases, $l$ is any positive integer, $\alpha$ and $\beta$ are
complex group parameters. An orthonormal basis in $\cal L$ is given by
functions $\phi_\nu=c_\nu z^\nu$, where $\nu$ is a nonnegative integer and
$c_\nu$ is a normalization constant. For the compact case, $S^2$,
the manifold has a finite volume, which equals the total number of states,
$V=N=l+1$. (The norm of $\phi_\nu$, Eq. (5), does not exist if $\nu>l$.)
For the pseudosphere, the Hilbert space is infinite, and the domain in
$\;\IC$ is noncompact, $z\bar{z}<1$. The monomials given above are always
eigen-functions of the Hermitean operator $\hat{H}=zd/dz$, which has the
integer spectrum, like the one-dimensional harmonic oscillator in the
Euclidean case. Thus we have got all the unitary representations of $SU(2)$
(corresponding to the angular momentum values $\frac{1}{2}l$), and the
representations of the discrete series for $SU(1,1)$.

The usual quantum mechanics in one degree of freedom is the boundary case
between $S^2$ and $H^2$. Actually, if the system motion is confined to a
restricted domain $|z|\ll 1$, one can introduce the ususal phase space
 and write $z=(q+ip)/\sqrt{2}R$, so that for $R\to\infty$ one gets
the Hilbert space of Eq. (1) for $m=1$, $\hbar=R^2/l$ and $l\gg 1$.
 Berezin noted that $\hbar^{-1}$ must have a discrete spectrum for $S^2$;
he remarked that this condition ``seems extravagant"\cite{ber75}.
In fact, the ``quantization" of the K\"{a}hler structures is typical;
it results from Eq. (29), if one has a nontrivial representation of a
compact subgroup in the group of motions.
Here the subgroup is $U(1)$: $\beta=0,\;\alpha=e^{it}$.

The quantization on sphere was proposed by Souriau\cite{souriau}.
However, a decade before, Klauder\cite{klaud60} was probably the first who
considered the quantization on sphere (actually, on the direct product of
infinitely many spheres) for the spinor representation
(the particular case of $l=1$ ), introduced the coherent states, and used the
method for quantum
field theory. Berezin \cite{ber75a} considered classical symmetric complex
domains, and the Lobachevsky plane was the simplest particular case.

\section{The Lie algebra and the Poisson brackets}

The group action in $\cal L$ is represented by the unitary operator
     \begin{equation}
\hat{U}(g)\psi(z)=\exp[-\Phi(z;g^{-1})]\psi(g^{-1}z),
   \end{equation}
which leads to a representation of the Lie algebra $\bf g$ in terms of the
holomorphic first-order differential operators,
     \begin{equation}
\tau_a\;\to\;\hat{T}_a=\nabla_a-\varphi_a(z),\;\;\;
[\hat{T}_a,\hat{T}_b]=f^c_{ab}\hat{T}_c,
   \end{equation}
cf. Eqs. (25) and (28).

The symbol representation for the operators is obtained from Eq. (11),
     \begin{eqnarray}
(\hat{U}(g)\psi)(\zeta)=\int_{\cal M}U_g(\zeta,\bar{z})\psi(z)
\exp\left[ K(\zeta,\bar{z})-K(z,\bar{z})\right]d\mu(z,\bar{z}),\\
U_g(\zeta,\bar{z})=\exp\left[ K(g^{-1}\zeta,\bar{z})-K(\zeta,\bar{z})
-\Phi(\zeta;g^{-1})\right]\\
\equiv\exp\left[K(\zeta,\overline{gz})
-K(\zeta,\bar{z})+\overline{\Phi^(z;g)}\right],\nonumber\\
T_a(\zeta,\bar{z})=\nabla_aK(\zeta,\bar{z})-\varphi_a(\zeta)\equiv
-\overline{T_a(z,\bar{\zeta})}=
-\ov{\nabla_a}K(\zeta,\bar{z})+\overline{\varphi_a(z)}.
   \end{eqnarray}
(The second equality is true for the real basis, where
$\hat{T}_a=-\hat{T}_a^\dagger$.) Thus the symbols of the Lie algebra
are expressed in terms of the Killing fields and the symplectic
one-form generated by the  K\"{a}hler structure,
\begin{equation}
T_a(\zeta,\bar{z})=\kappa(\zeta)^\alpha_a\Lambda_\alpha(\zeta,\bar{z})
-\varphi_a(\zeta),\;\;\;dK(\zeta,\bar{z})=
\Lambda_\alpha d\zeta^\alpha+\Lambda_{\bar{\beta}}d\bar{z}^\beta.
\end{equation}

Assuming that the (1,1)-form $\omega$ is non-degenerate, one can define
the Poisson brackets in $\cal M$ by means of a field $\varpi$ dual to $\omega$.
 Namely, for any two symbols $A(z,\bar{z})$ and $B(z,\bar{z})$ one has
       \begin{eqnarray}
\{A,B\}_{\rm P.B.}\equiv \varpi^{\alpha\bar{\beta}}(
\partial_\alpha A\partial_{\bar{\beta}}B-
\partial_\alpha B\partial_{\bar{\beta}}A)=-\{B,A\}_{\rm P.B.}\\
\omega_{\alpha\bar{\beta}}\varpi^{\alpha\bar{\gamma}}=
\delta_{\bar{\beta}}^{\bar{\gamma}},\;\;\;
\omega_{\alpha\bar{\beta}}\varpi^{\gamma\bar{\beta}}=
\delta_\alpha^\gamma.\nonumber
        \end{eqnarray}
The Jacobi identity results from the fact that $\omega$ in (9) is a closed
form. Using (35), one can show that the Poisson brackets provide with a
representation for the Lie algebra $\bf g$ ,
      \begin{equation}
\{T_a,T_b\}_{\rm P.B.}=\nabla_aT_b-\nabla_bT_a=f^c_{ab}T_c.
   \end{equation}

The comparison between Eqs. (32) and (38) shows the meaning of the
``quantization" performed. The Lie multiplication in $\bf g$,
Eqs. (25) and (32),  plays the same role as the {\em canonical
commutation relations} in the standard quantum theory: there is a one-to-one
correspondence between commutators and the Poisson brackets\cite{dirac}
for the {\em canonical variables} $T_a$.  Of course, the correspondence is
maintained necessarily only for the canonical variables, like it holds for
coordinates and momenta in the usual theory.

The coherent states (10) are solutions to the following differential
equations
     \begin{equation}
\hat{T}_a\Psi_v(z)=T_a(z,\bar{v})\Psi_v(z).
   \end{equation}
As in the usual quantum mechanics, they minimize the uncertainty relations
(cf. Ref.\cite{perelomov}).

\section{K\"{a}hler structures for compact Lie groups}

\subsection{Borel coordinates}

The K\"{a}hler potentials have been constructed explicitly for all compact
simple Lie groups\cite{bando,itoh,bordem,bmar}. The complex parameters are
introduced
following the Borel method\cite{borel}.

The following notations will be used.
Let $\cal G$ be a compact simple Lie group, and $\cal T$ be its maximal
Abelian subgroup (the maximal torus); the coset space $\cal F=G/T$ is
the flag manifold. The local complex parametrization in $\cal F$ is
introduced by means of the canonical diffeomorphism $\cal G/T\cong G^{\rm
c}/P$.
Here $\cal G^{\rm c}$ is the complex extension of $\cal G$;
the parabolic subgroup $\cal P$  is a semi-direct product of $\cal T$ and
the Borel subgroup ${\cal B\subset G}^{\rm c}$.
We shall employ the canonical basis $\{\tau_a\}=\{h_j,e_{\pm\alpha}\}$
in the Lie algebra $\bf g$, where $j=1,\cdots,r\equiv{\rm rank}({\bf g})$
and $\{\alpha\}\in\Delta^+_{\bf g}$ are the positive roots of ${\bf g}$
(see e.g. in Ref.\cite{bourbaki}).  The number of the positive roots is
$n_+=\half[{\rm dim}({\bf g})-{\rm rank}({\bf g})]$.
The Lie products of the basis elements, important for following, are
    \begin{equation}
[h_j,h_k]=0,\;\;\;[h_j,e_\alpha] =(\alpha\cdot{\bf w}_j)e_\alpha,\;\;\;
[e_\alpha,e_\beta]=\chi(\alpha,\beta)e_{\alpha+\beta}.
    \end{equation}
Here $\chi(\alpha,\beta)$ is a function on the root lattice, which vanishes
if $\alpha+\beta\not\in\Delta^+_{\bf g}$, and ${\bf w}_j$ are the fundamental
weight vectors. For any unitary irreducible group representation,
its dominant weight $\bf l$ is given by
a sum of the fundamental weights with nonnegative integer coefficients,
     \begin{equation}
{\bf l}=\sum^r_{j=1}l^j{\bf w}_j.
    \end{equation}

Given the canonical basis, the Lie algebra $\bf g^{\rm c}$ is splitted into
three subalgebras, $\bf g^{\rm c}=g^-\oplus t^{\rm c}\oplus g^+$,
corresponding to three subsets of the basis elements,
$\{e_{-\alpha}\}$, $\{h_j\}$, $\{e_{\alpha}\}$.
Respectively, the Lie algebra $\bf g^+$ generates a nilpotent subgroup
$\cal G^+\subset G^{\rm c}$, and the Lie algebra
${\bf p=g^-\oplus t}^{\rm c}$ generates $\cal P$.
The complex parameters which are introduced in $\cal F$
correspond to the positive roots of $\bf g$, and (complex) dim$({\cal F})=n_+$,
    \begin{equation}
f(z)=\exp(\sum_{\alpha\in\Delta^+_{\bf g}}z^\alpha e_\alpha)\in{\cal G}^+,
\;\;\;z^\alpha\in\IC.
    \end{equation}
Similarly, one can write elements of the parabolic subgroup.
Any element of the complex group has a unique Mackey decomposition,
    \begin{eqnarray}
\forall g\in {\cal G}^{\rm c},\;\;\;
g=fp,\;\;\;f\in {\cal G}^+,\;p\in {\cal P},\\
p=\exp(\sum_{j=1}^rx^jh_j+
\sum_{\alpha\in\Delta^+_{\bf g}}y^\alpha e_{-\alpha}).\nonumber
    \end{eqnarray}
(The decomposition is valid for all $g$, except for a subset of a lower
dimensionality). As soon as  $f(z)$ is an element of a nilpotent group,
its matrix representations are polynomials of $z^\alpha$.
 The local form (42) for $f$ is valid in a neighbourhood of the point
$z^\alpha=0$, i.e. the origin of the coordinate system in ${\cal F}$.
Transition to other domains of $\cal F$, covering the manifold completely,
can be performed by means of the group transformations.
In order to obtain the decomposition (43) for any given $g$,
one has first to calculate $f(z)^{-1}g=p$, using the fact that
$f(z)^{-1}\in{\cal G}^+$ is a polynomial of $z_\alpha$
in any group representation. Setting $p=\exp(\eta)$,
$\eta\in{\bf g}^{\rm c}$, one requires that $\eta$ has no
$e_\alpha$-components. Thus one gets $n_+$ algebraic equations for $n_+$
variables $z_\alpha$. As soon as the equations are solved and $z$'s are
found as functions of $g$, one gets $f(z)$ and $p=f^{-1}g$.

The group $\cal G^{\rm c}$ acts on $\cal F$ by left multiplications.
Actually, for any $g$ and $z$ one has a unique decomposition,
     \begin{equation}
gf(z)=f(gz)p(z;g),\;\;\;\;p(z;g)\in {\cal P},
    \end{equation}
where $gz$ is a {\em rational} function of $z$. For any element $g$ which
does not drive the point with coordinates $z^\alpha$ outside the coordinate
neighbourhood containing the origin where (42) is valid, $gz$ and $p(z;g)$
can be obtained from (44) by means of algebraic operations. Performing two
consecutive transformations, like in Eq. (23), one gets
     \begin{equation}
p(z;g_2g_1)=p(g_2z;g_1)p(z,g_2),\;\;\;\;
  \forall g_1,g_2\in {\cal G}^{\rm c}.        \end{equation}
The decomposition in Eq. (44) shows the way to the desired K\"{a}hler
structure.

\subsection{K\"{a}hler potentials}

In a work by Bando, Kuramoto, Maskawa and Uehara\cite{bando}
the K\"{a}hler potentials were expressed in terms of the fundamental
unitary representation of $\cal G$. The representation $g\to\hat{g}$
(hat stands for the matrix in this section) is also a representation
of ${\cal G}^{\rm c}$, but the elements of ${\cal G}^+$ and $\cal P$
are not unitary, if they do not belong to $\cal G$. The {\em partial}
solution of Eqs. (6), (7) is given in terms of the generalized
determinant (see Appendix),
      \begin{eqnarray}
K^j(\zeta,\bar{z})=
\log{\det}'\left(\hat{f}(\zeta)\theta_j\hat{f}(z)^\dagger\right),\\
\Phi^j(z;g)=-
\log{\det}'\left(\hat{g}\hat{f}(z)\theta_j\hat{f}(gz)^{-1}\right).
      \end{eqnarray}
Here  $\theta_j$ is a projection matrix which satisfies the following
conditions
     \begin{eqnarray}
\theta_j=\theta_j^\dagger ,\;\;\;\theta_j^2=\theta_j,\;\;\;
{\det}'\theta_j=1,\nonumber\\
\theta_j\hat{p}\theta_j=\hat{p}\theta_j,\;\;\;\forall p\in{\cal P}.
    \end{eqnarray}
There are $r$ independent matrices of this kind ($j=1,\cdots,r$),
having different ranks.
It is easy to see, using Eqs. (44) and (A4), that for each $\theta_j$
     \begin{eqnarray}
\frac{\det'[\hf(g\zeta)\theta\hf(gz)^\dagger]}{\det'\theta}=
\frac{\det'[\hat{g}\hf(\zeta)\hat{p}^{-1}\theta(\hat{p}^{-1})^\dagger
\hf(z)^\dagger\hat{g}^\dagger]}
{\det'[\hat{p}^{-1}\theta(\hat{p}^{-1})^\dagger]}
\frac{\det'\theta}{\det'[\hat{p}\theta\hat{p}^\dagger]}\nonumber\\
=\frac{\det'[\hf(\zeta)\theta\hf(z)^\dagger]}{\det'\theta}
\frac{\det'\theta}
{\det'[\theta\hat{p}(\zeta;g)\theta^2\hat{p}(z;g)^\dagger\theta]}.
   \end{eqnarray}
The second factor is a holomorphic function of $\zeta$ times
an anti-holomorphic function of $z$, owing to Eq. (A3).
One can see (cf. Eq. (50) below) that rank$(\theta_j)$ equals the number
of weights $\bf w$ in the fundamental representation, which satisfy the
condition $({\bf w\cdot w}_j)=({\bf w}_1\cdot {\bf w}_j)$.
(Here ${\bf w}_1$ is the dominant weight of the fundamental
representation, i.e. the lowest fundamental weight.) In Appendix B,
the ranks of $\theta_j$ are given for all the classical groups and
for the exceptional group $G_2$.

A matrix having the properties described above can be constructed for
any subgroup $U(1)\subset{\cal G}$, and the basis matrices $\theta_j$
correspond to the components of the maximal torus $\cal T$.
In order to see that, let us take a (noncompact) subgroup
$q_j(t)=\exp(-ith_j)\in{\cal G}^{\rm c}$, and consider the
transformation $p\to p_t=q_j(t)pq_j(t)^{-1}$. (Note that $\hat{h}_j$
is anti-self-adjoint for unitary representations.) It is easy
to see that $p_t\in{\cal P}$ and its parameters are $y^\alpha_t=
y^\alpha\exp[-t(\alpha\cdot{\bf w}_j)]$. As soon as
$(\alpha\cdot{\bf w}_j)\geq 0$ for all positive roots $\alpha$, only
those parameters $y^\beta$ survive in the limit $t\to\infty$, for which
$(\beta\cdot{\bf w}_j)= 0$. In the other words, $p_\infty$ is reduced to
a subgroup ${\cal P}_j$, generated by $e_{-\beta}$ for the roots $\beta$
satisfying the above condition. This observation suggests the following
construction of the projection matrices in terms of elements of the
Cartan subalgebra. For any unitary representation, where $\hat{h}_j$ is
diagonal, $\theta_j$ is also diagonal and has $1$ at the sites where
$-i\hat{h}_j$ has its maximum eigen-value, which is
$v_j\equiv({\bf w}_j\cdot{\bf w}_1)$. All the other elements of
$\theta_j$ are $0$. Thus the fundamental K\"{a}hler potentials can be
also written in the following invariant form
     \begin{equation}
K^j(\zeta,\bar{z})=\lim_{\lambda\to 0}\lim_{t\to\infty}\log
\frac{\det[\hf(\zeta)\hat{q}_j(t)\hf(z)^\dagger-\lambda\exp(v_jt)]}
{\det[\hat{q}_j(t)-\lambda\exp(v_jt)]}.
   \end{equation}
The order of the limits cannot be changed, of course.

Evidently, the sum of two K\"{a}hler potentials has also the desired
transformation property (6), so the fundamental potentials generate
a lattice. Ultimately, the {\em general} solution is given for
any group representation with a dominant weight $\bf l$,
  \begin{eqnarray}
K^{\bf (l)}(\zeta,\bar{z})=\sum^r_{j=1}l_jK^j(\zeta,\bar{z}),\\
\Phi^{\bf (l)}(z;g)=\sum^r_{j=1}l_j\Phi^j(z;g).
   \end{eqnarray}
The fundamental potentials (46) are logarithms of polynomials
in $\zeta$ and $z$, so the function $\exp [K^{\bf (l)}]$, which appears
in the integrals in Section 2, is a polynomial, and its
degree is determined by $\bf l$.

If any unitary representation of $\cal G$ would be employed in Eq. (46),
the result can be reduced to a sum of the fundamental potentials $K^j$,
as soon as that representation can be extracted from a direct product of
the fundamental ones.

Having the expression for the K\"{a}hler potential, one gets immediately
symbols for elements of the group $\cal G$ and its Lie algebra $\bf g$,
     \begin{eqnarray}
U_g^{\bf (l)}(\zeta,\bar{z})=
\prod^r_{j=1}\left[ \frac
{{\det}'\left(\hat{g}\hf(\zeta)\theta_j\hf(z)^\dagger\right)}
{{\det}'\left(\hf(\zeta)\theta_j\hf(z)^\dagger\right)}\right]^{l_j},\\
T_a^{\bf (l)}(\zeta,\bar{z})=
{\rm tr}\left(\Theta^{\bf (l)}(\zeta,\bar{z})\hat{\tau}_a\right).
    \end{eqnarray}
Here the trace is taken in the fundamental representation, and
the matrix $\Theta$, called sometimes {\em momentum map},
has been described in Ref.\cite{bmar}. It is a sum of the
fundamental components,
     \begin{equation}
\Theta^{\bf (l)}(\zeta,\bar{z})=
\sum^r_{j=1}l_j\Theta_j(\zeta,\bar{z}).
    \end{equation}
The fundamental momentum maps $\Theta_j(z,\bar{z})$ may be
considered as the projections $\theta_j$ transported from the origin
of $\cal M$ to its arbitrary point. In the other words, there exists
an element $v(\zeta,\bar{z})\in\cal G$, such that
     \begin{equation}
\Theta_j (\zeta,\bar{z})=\hat{v}\theta_j\hat{v}^{-1},\;\;\;
\hat{v}^{-1}=\hat{v}^\dagger.
    \end{equation}
It was shown\cite{bmar} that
$\Theta_j(\zeta,\bar{z})\exp[K^j(\zeta,\bar{z})]$ is a polynomial
in $\zeta$ and $z$. The Killing fields and the Lie algebra cocycles
$\varphi_a$, which are present in the general expression for $T_a$
in Eq. (36), have been also given in Ref.\cite{bmar}.

It is easy to see that $K^j$ satisfy the following differential
equations
      \begin{equation}
\nabla_\beta(\zeta) K^j(\zeta,\bar{z})-
\ov{\nabla_\beta (z)}K^j(\zeta,\bar{z})=0,\;\;\;
\forall\beta:\;(\beta\cdot{\bf w}_j)=0.
   \end{equation}
Therefore the (1,1)-form $\omega$, defined in Eq. (9),
is degenerate if there is a non-empty set of roots $\sigma$,
for which $K^{\bf (l)}$ satisfies Eqs. (57),
     \begin{equation}
({\bf l}\cdot\sigma)=0, \;\;\;\;
\sigma\in\Delta^+_{\bf s}\subset\Delta^+_{\bf g}.
   \end{equation}
Actually, the set $\Delta^+_{\bf s}$ is generated by the primitive
roots $\gamma_j$, for which $l_j=0$. Now the K\"{a}hler structure
is introduced in a subset of $\cal F$, given by the constraints
$z_\sigma=0$. The flag manifold $\cal F$ can be considered as
a fiber bundle, $\cal M$ being its base, where the unitary group
representation ${\sf R}_{\bf l}$ generates a non-degenerate K\"{a}hler
structure. The local coordinates on $\cal M$ are $z_\alpha$, where
$\alpha\in\Delta^+_{\bf g}\setminus\Delta^+_{\bf s}$. Respectively,
the little group of $\cal M$ is larger than the maximal torus,
     \begin{equation}
\cal M=G/H,\;\;\;\;H=S\otimes T',
     \end{equation}
where $\cal S$ is the semi-simple Lie group having $\bf s$ as its Lie
algebra, and $\cal T'\subset T$ is a torus generated by those basis
elements $h_j$, for which $l_j\neq 0$. This construction has a clear
interpretation in terms of Dynkin graphs\cite{bordem}.
Given a group representation ${\sf R}_{\bf l}$, one has to eliminate
from the Dynkin graph the nodes for which $l_j\neq 0$. The remaining
nodes indicate a semi-simple Lie algebra $\bf s\subset h\subset g$.
The parabolic subgroup $\cal P$ is extended respectively; its Lie
algebra is ${\bf p=g^-+t}'^{\rm c}+{\bf s}^{\rm c}$.

\subsection{Unitary groups}

Fot ${\cal G}=SU(r+1)$, the fundamental representation is
$(r+1)$-dimensional. The primitive roots and the fundamental
weights are given in terms of an orthonormal basis $\{\epsilon_i\}$
in  the Euclidean space $\;\IR^{r+1}$ (see e.g. in Ref.\cite{bourbaki})
   \begin{equation}
\gamma_j=\epsilon_j-\epsilon_{j+1},\;\;\;{\bf w}_j=
\sum_{i=1}^j\epsilon_i-\frac{j}{r+1}\sum_{i=1}^{r+1}\epsilon_i.
    \end{equation}
Now $v_j=1-j/(r+1)$, and rank$(\theta_j)=j$.
The local coordinates corresponding to the positive roots
$\alpha=\epsilon_j-\epsilon_k$ are elements of a triangular matrix
$\hat{z}$, i.e. $z_{jk}$, $1\leq j < k\leq (r+1)$
(other elements of the matrix are zero), and the complex dimensionality
of $\cal F$ is $\frac{1}{2}r(r+1)$. The matrix $\hf (z)$ is triangular;
its diagonal elements are $1$, and polynomials in $z^\alpha$ stand
above the diagonal. (The general form of the polynomials can be
written down easily.) The manifold $\cal F$ has an additional symmetry
under a reflection of the root space, under which $j\rightarrow r-j$,
and $\theta_{r-j}\rightarrow \upsilon(I-\theta_j)\upsilon^{-1}$,
where $\upsilon$ is a matrix reversing the order of components in the
representation space, corresponding to the uatomorphism of the root
system, specific for the unitary groups.

If the number of nonzero components $l_j$ is $k<r$, the phase space is a
section of the flag manifold, $\cal M\subset F$.
Now the gauge group is larger than ${\cal T}_r$,
namely ${\cal H=T}_k\otimes\prod SU(r_i+1)$, where $\sum r_i=r-k$.
The corresponding Dynkin graph is obtained by elimination of $k$
links from the chain describing $SU(r+1)$. If only one component
is nonzero, say, $l_q=l$, one gets the Grassmann manifold
${\cal M}={\sf Gr}(p,q)\equiv U(p+q)/U(p)\otimes U(q)$ (where
$1\leq q\leq p<r+1\equiv p+q$) which is a rank-one section of $\cal F$.
The complex dimensionality of the manifold is $pq$, and the local
coordinates are elements of a $p\times q$ matrix $\hat{z}$, so that the
elements of $\hf$ are $f_{jn}=\delta_{jn}+z_{j,n-p}$, where $1\leq j\leq p$,
and $p+1\leq n\leq p+q$. The resulting K\"{a}hler potential is
        \begin{equation}
K(z,\bar{z})=l\log\det(I_q+\hat{z}^\dagger\hat{z}).
         \end{equation}
For $q=1$, $\hat{z}$ is a complex vector, ${\cal M}\equiv{\sf CP}^r$ is
the complex projective space, and one gets the familiar Fubini -- Study
metrics\cite{nomizu}. The metrics is ``quantized", since $l$ is integer.

Extending the arguments given in Section 2.4, one can claim that if $l$
is large it evaluates the area of two-dimensional cross sections of the
phase space $\cal M$ in the $\hbar$ units, i.e. $\hbar\sim l^{-1}$.
If the rank of $\cal M$ is $k>1$, one has a number of nonzero numbers
$l_j$. Introducing a local system of coordinates and momenta, one gets
an apparent anisotropy in $\hbar$, which cannot be eliminated because of
the boundary conditions.

\section{Dynamics}

Until this section, we have been discussing {\em quantum kinematics}
(as defined in Schwinger's book\cite{schwin70}). In order to introduce
{\em dynamics}, one needs a Hamiltonian $\hat{H}$, to be given as a
function of the canonical variables $\hat{T}_a$, like Hamiltonians of
the standard theory are expressed in terms of coordinates and momenta.
Thus $\hat{H}$ belongs to the universal envelopping algebra of
$\bf g$. The problem is to get the evolution operator
$\hat{G}(t)=\exp(-it\hat{H})$, where $t$ stands for {\em time},
and to find solutions to the Heisenberg and/or Schr\"{o}dinger
equations of motion,
     \begin{eqnarray}
d\hat{T}_a/dt=-i[\hat{T}_a,\hat{H}],\;\;\;
\hat{T}_a(t)=\hat{G}(t)^{-1}\hat{T}_a(0)\hat{G}(t);\\
d\hat{\rho}/dt=i[\hat{\rho},\hat{H}],\;\;\;
\hat{\rho}(t)=\hat{G}(t)\hat{\rho}(0)\hat{G}(t)^{-1},
    \end{eqnarray}
where $\hat{\rho}$ is the system density operator. In the functional
approach, one has to calculate the partition function $Z(t)$ and the
generating functional $F_\eta$, which shows the system response to
external (time-dependent) {\em source terms} in the Hamiltonian,
      \begin{eqnarray}
i\partial\hat{G}_\eta/\partial t=
[\hat{H}+i\eta^a(t)\hat{T}_a]\hat{G}_\eta,
\;\;\;\hat{G}_\eta(t_0,t_0)=\hat{I},\\
F_\eta=\lim_{t_\pm\to\pm\infty}{\rm Tr}\;\hat{G}_\eta(t_+,t_-),
\;\;\;Z(t)={\rm Tr}\;\hat{G}_0(t_0+t,t_0).
      \end{eqnarray}
If the symbols are used, the equations of motion are linear
integro-differential equations, and the trace is an integral in $\cal M$.
In constructing possible Hamiltonians as functions of $\hat{T}_a$, one
should have in mind that for compact spaces the group representations
are finite-dimensional, and there are some identities for any given
representation. In particular, each Casimir operator $\hat{C}_n$ is
the unit operator times a number, which depends on $\bf l$.

The solution is obtained immediately for Hamiltonians which are
elements of $\bf g$ combined with the Casimir operators,
i.e. $\hat{H}=\sum\mu_n\hat{C}_n+i\xi^a\hat{T}_a$, where $\mu_n$ and
$\xi^a$ are real, as soon as $\hat{H}$ is self-adjoint.
In this case, the evolution operator is just an element of
$\cal G$, its symbol is given by Eq. (53), and $Z(t)$ is the
representation character, times $\exp(-it\sum\mu_n\lambda_n)$,
where $\lambda_n$ are eigen-values of $\hat{C}_n$.

Let us consider, for instance, the Hamiltonian
$\hat{H}=\mu\hat{C}_2+i\xi^a\hat{T}_a$ for sphere
$S^2\equiv SU(2)/U(1)$. Its symbol is
    \begin{equation}
H(z,\bar{z})=\frac{\mu}{4}l(l+2)+l(1+z\bar{z})^{-1}
[\xi^1(z+\bar{z})-i\xi^2(z-\bar{z})+\xi^3(1-z\bar{z})].
   \end{equation}
By means of a change of coordinates (rotation) one can set
$\xi^1=0=\xi^2$, and the spectrum is obtained immediately,
$\varepsilon_\nu={\rm const}+\xi\nu$, where $\xi=|\xi^3|$ and
$\nu=0,1,\cdots,l$. The constant $\mu$ may be chosen to set
$\varepsilon_0=0$. The partition function is
    \begin{equation}
Z(t)=\exp(-il\xi t)\frac{\sin (l+1)\xi t}{\sin\xi t}.
   \end{equation}
The classical trajectories are the parallels of latitude
on the sphere, namely,
$z(t)=z(0)e^{-i\xi t}$. In the limit, discussed in section 2.4,
where $l\to\infty$ and $z=(q+ip)/\sqrt{2l\hbar}$,
we get the standard Hamiltonian of the one-dimensional oscillator,
retaining terms of the order of $z\bar{z}$,
i.e. $\hbar H(p,q)=\xi(p^2+q^2)+{\rm const}$.

The semi-classical approximation works usually in the
situations where large parameters are present in exponential
integrands, and the integrals are evaluated by means of the steepest
descent method. This is true as well for dynamics on homogeneous
K\"{a}hler manifolds. The large parameter appears if some
components of ${\bf l}$ are large. By the way, the total number of
states, which is proportional to the volume of $\cal M$, is also large
in this case. The large parameter must be present also in the
Hamiltonian $H$, otherwise the classical dynamics would be trivial.
In order to get the semi-classical approximation, let us construct
an analogue of the path integral. The standard method is to use
the identity
 \begin{equation}
\hat{U}_t\equiv  e^{-i\hat{H}t}=
\left( e^{-i\hat{H}t/{\cal N}}\right)^{\cal N},\;\;\;\;\forall{\cal N},
     \end{equation}
and write an approximate expression for the symbol of the evolution
operator at small times, $\tau=t/{\cal N}$,
where ${\cal N}\rightarrow\infty$,
 \begin{equation}
e^{-i\tau\hat{H}}\rightarrow e^{-i\tau H(z,\bar{z})}+O(\tau^2).
         \end{equation}
In principle, this is not correct for our manifolds, because each symbol
must be a rational function, i.e. a polynomial divided by $\exp(K)$, and
this form does not survive the exponentiation. In the limit of large
$\bf l$, however, the polynomials are of a very high degree, and
this condition is not too restrictive. Multiplying $\cal N$ operators
(69) and calculating the trace (cf. eq. (20)), one gets the partition
function
    \begin{eqnarray}
Z(t)=\lim_{{\cal N}\rightarrow\infty}\int_{\cal M}\!\cdots\!\int
    \prod_{n=1}^{\cal N}d\mu(z_n,\bar{z}_n)\nonumber\\
\exp\left\{\sum_{n=1}^{\cal N}\left[K(z_n,\bar{z}_{n+1})-
      K(z_n,\bar{z}_n)-i\tau H(z_n,\bar{z}_{n+1})\right]\right\}.
  \end{eqnarray}
The boundary condition is $\bar{z}_{{\cal N}+1}\equiv\bar{z}_1$,
so the sequence of points on $\cal M$ can be considered as closed.
The exponent has an extremum under the following conditions,
for $n=1,\ldots,{\cal N}$,
 \begin{eqnarray}
\Lambda_\alpha(z_n,\bar{z}_{n+1})-\Lambda_\alpha(z_n,\bar{z}_n)-
i\tau\partial_\alpha H(z_n,\bar{z}_{n+1})=0,\nonumber\\
\bar{\Lambda}_\alpha(z_{n-1},\bar{z}_n)-
\bar{\Lambda}_\alpha(z_n,\bar{z}_n)-
i\tau\partial_{\bar{\alpha}} H(z_{n-1},\bar{z}_n)=0,
       \end{eqnarray}
where $\Lambda$ and $\bar{\Lambda}$ are the partial derivatives
of $K$, as given in Eq. (36). Under reasonable conditions on
the Hamiltonian, in the limit $\tau\rightarrow 0$, the region
$z_{n+1}\rightarrow z_n$
contributes predominantly to the integral, so that
        \begin{equation}
\Lambda^\alpha(z_n,\bar{z}_{n+1})-\Lambda^\alpha(z_n,\bar{z}_n)\approx
\omega^{\alpha\bar{\beta}}(z_n,\bar{z}_n)(\bar{z}_{n+1}-\bar{z}_n)_\beta,
       \end{equation}
and equations (71) become the usual Hamilton equations of motion,
which can be set to the familiar form with the Poisson brackets
defined in Eq. (37),
        \begin{equation}
\frac{dz^\alpha}{dt}=\{H,z^\alpha\}_{\rm P.B.},\;\;\;
\frac{d\bar{z}^\alpha}{dt}=\{H,\bar{z}^\alpha\}_{\rm P.B.}\;\;.
       \end{equation}
These equations of motion can be also derived from the variational
principle applied to the action which results from the sum in Eq. (70)
if the difference of two K\"{a}hler potentials is replaced by the
differential,
       \begin{equation}
{\cal A}_t=\int^t_0\{\half i[\Lambda^\alpha(z,\bar{z})dz^\alpha-
\bar{\Lambda}^\alpha(z,\bar{z})d\bar{z}^\alpha]
-H(z,\bar{z})d\tau\}.
        \end{equation}
(We added the total derivative $\half dK$ to the integrand to make the
action real. Note that $K$ is not defined globally on $\cal M$,
but the addition is possible because of the cocycle condition.)
The action functional is calculated on closed trajectories, $z(0)=z(t)$.

The arguments presented above support the meaning of the construction
as a method of quantization: the quantum theory has its classical limit
described by the action functional with the K\"{a}hlerian symplectic form.
The classical theory is local and does not require integer coefficients
at the fundamental  K\"{a}hler potentials.

\section{Concluding remarks}

Berezin's method results in the following construction.
For any unitary representation ${\sf R}_{\bf l}$ of a compact
simple group $\cal G$, one has a compact homogeneous
K\"{a}hler manifold $\cal M\equiv G/H$ and a Hilbert space $\cal L$
of (locally) holomorphic functions which can be considered as a line
bundle upon $\cal M$. According to the Borel -- Weil -- Bott theorem,
the Hilbert-space representation is unitary. The Lie algebra $\bf g$
is realized by means of linear differential operators in $\cal L$,
or by means of Poisson brackets for functions on $\cal M$.
This is the essence of {\em quantization}.
The K\"{a}hler potentials have been constructed explicitly for all
compact Lie groups.

Until now, the Berezin quantization had no real {\em physical
applications}, except for some ``model building". An extension
to a new theory of quantized fields will probably be the next
step.

We conclude with a few remarks on subjects which have been beyond
the scope of this paper.

{\em 1. Non-compact groups.} The method can be extended easily to
non-compact (locally-compact) groups which are subgroups of
${\cal G}^{\rm c}$. Let $\eta$ be a non-degenerate matrix in the
fundamental representation space of $\cal G$. A subgroup
$\tilde{\cal G}\in{\cal G}^{\rm c}$ can be specified with a
pseudo-unitarity condition $\hat{g}\eta\hat{g}^\dagger=\eta$.
The group $\tilde{\cal G}$ is non-compact if $\eta$ is not
positive-definite. The arguments of Section 4 are valid, if instead
of Eq. (46) the fundamental potentials are defined by
    \begin{equation}
K^j(\zeta,\bar{z})=
\log{\det}'\left(\hat{f}(\zeta)\theta_j\hat{f}(z)^\dagger\eta\right)
\end{equation}
(it is assumed that $\det'(\theta_j\eta)=1$). The pseudosphere in
section 2.4 is obtained in this way from $SU(1,1)\subset SL(2,\IC)$
with $\eta={\rm diag}(1,-1)$. More information on non-compact
homogeneous K\"ahler manifolds can be found elsewhere\cite{thesis}.
The manifold is non-compact, as it is an image of a bounded domain
in $\;\IC^m$ where $\det'$ in Eq. (75) is positive. The boundary of
the domain is an analogue of the absolute for the Lobachevsky plane.
The K\"ahler manifolds appear for discrete series of unitary
(infinite-dimensional) representations of non-compact groups.
In general, homogeneous manifolds of
Lie groups cannot be provided with a K\"{a}hler structure, their
geometry is too complicated (see e.g. in Ref.\cite{gorbat}).
In particular, it is impossible for $\tilde{\cal G}/\tilde{\cal H}$,
if $\tilde{\cal H}$ is non-compact.

{\em 2. Infinite-dimensional limit.} The main building blocks of the
method admit an extension to the limit of infinite-dimensional groups.
For example, infinite-dimensional Hilbert -- Schmidt Grassmannians
have been considered \cite{thesis}. It was found that the classical
Hamiltonians generate a Lie subalgebra of the total Poisson-bracket
Lie algebra, isomorphic to the central extension of $\bf g$ which
include the (quantum) anomaly. The geometric quantization in the
$N\rightarrow\infty$ limit has been also considered in a recent
paper\cite{martin}.

{\em 3. Super-manifolds.} For the compact case, representations of
the Lie groups can be constructed also in super-manifolds, i.e. in
the Grassmann algebra with anti-commuting generators (the exterior
algebra\cite{schwin70}). An example of $SO(3)$ was considered and
applied to the description of spinning paricles\cite{bermar}. Different
group representations are obtained on super-manifolds with different
numbers of Fermi-type degrees of freedom. Two approaches, based upon
the K\"ahler manifolds and the super-manifolds, are equivalent but
quite different technically. In field theory the equivalence is known
as ``bosonization of fermion models".

{\em 4. Integrals on K\"ahler manifolds.} Starting from the known
transformation properties, one can calculate a class of integrals
on $\cal M$, like Eq. (19).  A particular result is Weyl's
formula for the representation dimensionality and a similar formula
for volumes of the manifolds. The Weyl group can be also realized
in $\cal M$. (The result will be published elsewhere.)

{\em 5. Integrable systems.} Investigation of Lie subalgebras of
the universal envelopping algebra of $\bf g$ may provide with a new
insight into the theory of integrable systems (cf. the example in
Section 5).

{\em 6. Additional references.} Only a part of a great number of works
dealing with the Berezin approach to quantization has been mentioned
here. Besides those mentioned, see, for instance, Refs.
\cite{moreno,jmp,cmp,jgp}, and references therein.

{\em Acknowledgements}.
This work was has been completed when the second author was visiting
at the University of Texas at Austin. It is a pleasure to thank
C\'{e}cile DeWitt-Morette and other colleagues at the Center for
Relativity and the Department of Physics for their kind hospitality.
The support to the research from the NSF grant PHY9120042, G. I. F.
and the Technion V. P. R. Fund is gratefully acknowledged.

\section*{Appendix}

\subsection*{A. Generalized determinant}

For any self-adjoint positive semi-definite linear operator $A$,
its generalized determinant, det$' A$, is defined as the product
of all its {\em non-zero} eigen-values. Given a pair of semi-definite
operators of the same rank, $A$ and $B$, one has
     \begin{equation}
\frac{\det'A}{\det'B}=\lim_{\lambda\to 0}
\frac{\det (A-\lambda)}{\det (B-\lambda)}.
     \end{equation}
Applying the identity
    \begin{equation}
\log\frac{\alpha}{\beta}=-\int_0^\infty (e^{-t\alpha}-e^{-t\beta})
\frac{dt}{t},\;\;\;\forall \alpha,\beta > 0,
    \end{equation}
to the eigen-values of $A$ and $B$, and summung up, one gets
    \begin{equation}
\log\frac{\det'A}{\det'B}=-\int_0^\infty \left[
{\rm Tr}(e^{-tA})-{\rm Tr}(e^{-tB})\right]
\frac{dt}{t}.
    \end{equation}
This equality enables one to define the generalized determinants for elliptic
operators having definite partition functions;
it has been used in quantum field theory\cite{schwarz}.

We declare that two operators belong to the same class, and write
$A\sim B$, if there exists a non-singular operator $P$, relating them,
$B=PAP^\dagger$. One can show that for any non-singular operator $F$,
     \begin{equation}
\frac{\det'FAF^\dagger}{\det'A}=
\frac{\det'FBF^\dagger}{\det'B},
\;\;\;\;\;{\rm if}\;\; A\sim B.
     \end{equation}
Thus the ratio depends only on the class containing $A$ and $B$.
One can also show that
     \begin{equation}
\frac{\det'F_2F_1AF_1^\dagger F_2^\dagger}{\det'A}=
\frac{\det'F_1AF_1^\dagger}{\det'A}
\frac{\det'F_2AF_2^\dagger}{\det'A}.
     \end{equation}
This equality is an extension of the usual one,
$\det (F_1F_2)=\det F_1\det F_2$.

\subsection*{B. Projection matrices for classical groups and $G_2$}

We shall use the notations of Ref.\cite{bourbaki}. Roots and weights for
the simple Lie algebras of rank $r$ are given in terms of the orthonormal
basis $\{\epsilon_j\}$ in the root space, except for $A_r$ and $G_2$,
where the root space is described as a $r$-dimensional hyperplane in the
$(r+1)$-dimensional Euclidean space (the root space is normal to the vector
$\varepsilon=\sum_{j=1}^{r+1}\epsilon_j$). The primitive roots $\gamma_j$ are
   \begin{equation}  
        \begin{array}{rlr}
A_r: & \epsilon_j-\epsilon_{j+1} &  (1\leq j\leq r); \\
B_r: & \epsilon_j-\epsilon_{j+1},\; \epsilon_r & (1\leq j < r); \\
C_r: & \epsilon_j-\epsilon_{j+1},\; 2\epsilon_r & (1\leq j < r); \\
D_r: & \epsilon_j-\epsilon_{j+1}, \;\epsilon_{r-1}+\epsilon_r &
(1\leq j < r); \\
G_2: & \epsilon_1-2\epsilon_2+\epsilon_3, \;\epsilon_2-\epsilon_3. &
       \end{array}
  \end{equation}
The corresponding fundamental weights ${\bf w}_j$
are (the lowest vector ${\bf w}_1$ is the last in the line)
   \begin{equation}  
   \begin{array}{rlr}
A_r: & \sum_{i=1}^j\epsilon_i-\frac{j}{r+1}\varepsilon & (1\leq j\leq r); \\
B_r: & \sum_{i=1}^j\epsilon_i,\; \frac{1}{2}(\epsilon_1+\cdots +\epsilon_r) &
(1\leq j\leq r-1); \\
C_r: & \sum_{i=1}^j\epsilon_i & (1\leq j\leq r); \\
D_r: & \sum_{i=1}^j\epsilon_i,\; \frac{1}{2}(\epsilon_1+\cdots +\epsilon_r),\;
\frac{1}{2}(\epsilon_1+\cdots -\epsilon_r) & (1\leq j\leq r-2); \\
G_2: & 2\epsilon_1-\epsilon_2-\epsilon_3, \;\epsilon_1-\epsilon_3. &
  \end{array}
  \end{equation}
The weights of the fundamental representation, having ${\bf w}_1$ as its
dominant weight, and its dimensionality $d_f$ are given by
   \begin{equation}
   \begin{array}{rll}
A_r: & \epsilon_j-\frac{1}{r+1}\varepsilon,       & d_f=r+1; \\
B_r: & \frac{1}{2}(\pm \epsilon_1\pm \epsilon_2\cdots\pm \epsilon_r), &
d_f=2^r;\\
C_r: & (\pm \epsilon_1\pm \epsilon_2\cdots\pm \epsilon_r), & d_f=2^r;\\
D_r: & \frac{1}{2}(\pm \epsilon_1\pm \epsilon_2\cdots\pm \epsilon_r), &
d_f=2^{r-1};\\
G_2: & \pm (\epsilon_1-\epsilon_3),\pm (\epsilon_1-\epsilon_2),
\pm (\epsilon_2-\epsilon_3),\; {\bf 0},  & d_f=7.
  \end{array}
  \end{equation}
(For $D_r$ the number of minuses in each weight vector is odd.)

Ultimately, the ranks of the projection matrices $\theta_j$, corresponding
to the fundamental weights ${\bf w}_j$, are given in the following Dynkin
graphs. The rank of $\theta_1$, corresponding to ${\bf w}_1$, is 1.
\begin{equation}  
\begin{array}{rllcr}
                    & r        & r-1              & 2 & 1 \\
SU(r+1)\sim A_r:\;\;&\bigcirc\;\;-&\bigcirc-\cdots-&\bigcirc &-\;\;\bigcirc\\
                    &         &                   &            &       \\
                    & 2^{r-1} & 2^{r-2}           & 2       &  1 \\
SO(2r+1)\sim B_r:\;\;&\bigcirc\;\;-&\bigcirc-\cdots-&\bigcirc &=\!\!>\bigcirc\\
                    &         &                   &            &       \\
                    & 2^{r-1} & 2^{r-2}           & 2       &  1 \\
  Sp(2r)\sim C_r:\;\;&\bigcirc\;\;-&\bigcirc-\cdots-&\bigcirc &<\!\!=\bigcirc\\
                    &          &                  &            &       \\
                    & 2^{r-2}  & 2^{r-3}          & 2          & 1     \\
 SO(2r)\sim D_r:\;\;&\bigcirc\;\;- &\bigcirc-\cdots-&\bigcirc &-\;\;\bigcirc\\
                   &           &                  &\mid    &       \\
           &  &\;\;\;\;\;\;\;\;\;\;\;\;\;\;r&\bigcirc &     \\
                   & 2         & 1        & &       \\
G_2:               &\bigcirc\equiv\!\!>&\bigcirc &  &
\end{array}
\end{equation}
Similarly, one can calculate the ranks of $\theta_j$ for the other 4
exceptional groups. It is remarkable that all the projection matrices
can be constructed recursively.

\end{document}